\documentstyle[12pt,aaspp4]{article}
\def\ab{$\sim$}
\def\kms{km~s${}^{-1}$}

\def\Ks{{\it K$_{s}$}}
\def\K{{\it K}}
\def\H{{\it H}}
\def\J{{\it J}}
\def\Ls{{\it L$_{s}$}}

\lefthead{put in names for page header}
\righthead{put in abbreviated title for page header}

\begin{document}
\vskip 0.5in
\title{High Resolution Infrared Imaging of the Compact Nuclear Source in NGC4258}
\author{R. Chary\altaffilmark{1,5}, E. E. Becklin\altaffilmark{1}, 
A. S. Evans\altaffilmark{2}, 
G. Neugebauer\altaffilmark{3}, N. Z. Scoville\altaffilmark{2},
K. Matthews\altaffilmark{3}, M. E. Ressler\altaffilmark{4}}
\altaffiltext{1}{Division of Astronomy \& Astrophysics, 
University of California, Los Angeles, CA 90095-1562}
\altaffiltext{2}{Owens Valley Radio Observatory, California Institute of Technology, 
Pasadena, CA 91125}
\altaffiltext{3}{Palomar Observatory, California Institute of Technology, Pasadena, CA 91125}
\altaffiltext{4}{Jet Propulsion Laboratory, Pasadena, CA 91109}
\altaffiltext{5}{Present Address: Department of Astronomy \& Astrophysics, University of
California, Santa Cruz, CA 95064; rchary@ucolick.org}

\vskip 1in
\begin{abstract}

We present high resolution imaging of the nucleus of NGC4258 
from 1~$\micron$ to 18~$\micron$.
Our observations reveal that
the previously discovered compact source of emission 
is unresolved even at the near-infrared 
resolution of \ab0.2$\arcsec$ FWHM which
corresponds to about 7 pc at the distance of the galaxy. 
This is consistent with the source of emission
being the region in the neighborhood of the 
purported 3.5$\times$10$^{7}$ M$_{\sun}$ black hole.
After correcting for about 18 mags of visual extinction, the infrared data
are consistent with a F$_{\nu}\propto\nu^{-1.4\pm0.1}$ spectrum from
1.1~$\micron$ to 18~$\micron$, implying a non-thermal
origin.
Based on this spectrum, the 
total extinction corrected infrared luminosity (1$-$20~$\micron$)
of the central source is
2$\times$10$^{8}$~L$_{\sun}$. 
We argue that the
infrared spectrum and luminosity of the central source
obviates the need for a substantial contribution from
a standard, thin accretion disk at these wavelengths
and calculate 
the accretion rate through an advection dominated
accretion flow to be $\dot{M}\sim10^{-3}$~M$_{\sun}$/yr. 
The agreement between these observations and the theoretical spectral 
energy distribution
for advection dominated flows provides evidence for the 
existence
of an advection dominated flow in this low luminosity AGN.

\end{abstract}


\clearpage

\section{Introduction}

NGC4258 (M106) is a weakly active Seyfert 1 (Filippenko 1996) galaxy at a 
distance of \ab7 Mpc 
(we adopt H$_{0}$=65 km s$^{-1}$ Mpc$^{-1}$; at this distance
1$\arcsec$=33.9 pc). 
The nucleus of the galaxy is a prototypical low luminosity AGN.
Dynamical evidence for the existence of a massive black hole in the
nucleus comes from high spatial resolution VLBI observations of water
masers, which 
reveal the existence of a thin edge-on disk of inner radius 4 milliarcseconds
and outer radius 8 milliarcseconds (Miyoshi et al. 1995). The high velocity
features of the masers were found to be 
symmetrically offset from the systemic velocity by 700$-$1000 \kms~
as would be expected in a Keplerian disk. The inferred binding
mass of 3.5$\pm$0.1$\times$10$^{7}$ M$_{\sun}$ within a radius of 0.14 pc 
(Herrnstein, Greenhill, \& Moran 1996) makes
it the strongest supermassive black hole candidate after the nucleus of 
the Galaxy (Genzel et al. 1997, Ghez et al. 1998). The adopted binding mass
and distance for the galaxy are consistent with the more recent measurements 
of Herrnstein et al. (1999). Observations
by the X-ray satellite ASCA 
indicate the existence of a strongly obscured source 
with a photon index of $\Gamma$=1.8$\pm$0.3 
at the nucleus of the galaxy (Makishima et al. 1994). 
The photon index $\Gamma$=1+$\gamma$ where $\gamma$ is the spectral index.
The line of sight column density of
N$_{H}$=1.5$\times$10$^{23}$ cm$^{-2}$ derived from a fit to the
absorbed X-ray data, corresponds to A$_{V}$\ab80 mags
which would obscure the nucleus at visible wavelengths. 
The extinction-corrected X-ray luminosity from 2$-$10 keV is
4$\times$10$^{40}$~ergs~s$^{-1}$ and is relatively insensitive to
the absorption column density.
Based on arguments that the X-ray luminosity is typically about 5\% to
40\% of the bolometric luminosity in AGN (Mushotzky, Done, \& Pounds 1993), 
L$_{Bol}$ for the central engine in NGC4258 is of order
10$^{41}$~ergs~s$^{-1}$. This is \ab2$\times$10$^{-5}$~L$_{E}$, where L$_{E}$ is
the Eddington luminosity of the massive black hole at the nucleus and
corresponds to 4.4$\times$10$^{45}$~ergs~s$^{-1}$. 

The low bolometric luminosity of the central engine can be
explained by an incorrect
estimation of L$_{Bol}$ since the overall spectral energy distribution of
the central engine is not well known. Alternatively, since
L$_{Bol}$=$\eta\dot{M}c^{2}$ where $\eta$ is the radiative efficiency
and $\dot{M}$ is the mass accretion rate of the central engine, a low bolometric
luminosity could be attributed to:\\
1. a sub-Eddington mass accretion rate through a
standard, geometrically thin disk.
For the black hole in NGC4258, the Eddington mass accretion 
rate $\dot{M}_{E}$\ab0.75 M$_{\sun}$/yr where $\eta$ has
a fiducial value of 10\%.\\
2. a low radiative efficiency for the accretion phenomenon whereby most of the viscously
generated energy is carried through the event horizon of the black hole. This is more commonly called
an advection dominated accretion flow (ADAF; Narayan \& Yi 1995).

Strong line
of sight extinction to the nucleus of NGC4258, makes estimating the intrinsic
spectral energy distribution difficult.
Wilkes et al. (1995; hereafter W95)
detected a polarized continuum with F$_{\nu}\propto\nu^{-1.1\pm0.2}$ 
in scattered light
at visible wavelengths. The luminosity of the central source is only weakly constrained by
these observations to lie in the range 10$^{41}-10^{43}$ ergs s$^{-1}$
and is dependent on the scattering model.
Infrared observations are less susceptible to extinction effects and are
therefore a better probe of the energy spectrum and the output luminosity.
Chary \& Becklin (1997, hereafter CB97) detected a compact infrared source in the nucleus
of the galaxy. Based on the visible light polarimetric observations
of W95, CB97 concluded that the
intrinsic spectrum of the infrared
source is a power law (F$_{\nu}\propto \nu^{-1}$) and that the observed
\J$-$\H, \H$-$\K~colors could be explained by about
17 mags of visual extinction along the line of sight.
While the absolute extinction corrected luminosity at these wavelengths
is consistent with an ADAF model (Lasota et al. 1996), the derived spectrum
poses some problems because at these wavelengths,
both the ADAF model and the standard thin disk model
predict a $\nu^{\frac{1}{3}}$
Comptonized blackbody spectrum arising
from a cool geometrically thin, optically thick disk (Pringle 1981).
In addition, infrared emission from thermal reprocessing
of the central visible/UV continuum by dust in a flared disk or `torus' 
could not be unambiguously ruled out.
The only observation at mid-infrared wavelengths
which can potentially trace the dust distribution
or constrain the overall spectrum of the central source
is the 8$-$13~$\micron$ 5.7$\arcsec$ diameter data from Rieke \&
Lebofsky (1978; hereafter RL78). This measurement 
could be contaminated by extended
emission from regions of nuclear star formation.

Lasota et al. (1996) applied an ADAF model to fit the X-ray
spectrum of NGC4258. 
The model consists of an inner, optically
thin flow which is geometrically thick. The ADAF exists from an inner
radius of 3 R$_{S}$ 
up to a radius of 10$-$100 R$_{S}$ where 
R$_{S}$ is the Schwarzschild radius of the central black hole and
corresponds to 3.4$\times$10$^{-6}$ pc. The ions which are
heated by the viscously generated energy have a 
temperature of \ab10$^{12}$~K but their radiative timescale is longer than
the accretion timescale resulting in most of the energy being
transported into
the event horizon of the black hole by radial convection or `advection'.
The electrons (T$_{e}$\ab10$^{9}$ K) 
produce synchrotron photons which could be the seeds for 
self-Comptonization to produce a hard X-ray spectrum.
The model also includes a cooler geometrically thin, optically thick
disk which extends up to 10$^{6}$ R$_{S}$ to reproduce the blackbody bump 
seen at visible/UV wavelengths in most AGN.
This kind of accretion flow allows a `normal'
accretion rate of $\dot{m}$\ab0.2$\alpha$ but requires a
low radiative efficiency since most of the viscously generated energy is 
transported into
the event horizon of the black hole by the advection process. 
Here, $\alpha\leq 1$ is the
dimensionless viscosity parameter defined by Shakura \& Sunyaev (1973),
while $\dot{m}=\dot{M}(M_{\sun}/yr)/\dot{M_{E}}$ is the accretion rate
in units of the Eddington rate. Mahadevan (1997; hereafter M97) has 
investigated the spectral energy distribution from an ADAF in detail and shown that 
the Comptonization of soft cyclosynchrotron photons
results in a power law spectrum between 10$^{12}$ and 10$^{17}$ Hz.
The spectral index of the power law is essentially
constrained by the mass accretion rate through the ADAF.

In this paper, we present high resolution imaging of the nucleus of
NGC4258 at mid-infrared and near-infrared wavelengths.
The observations place a better handle on the spectrum and
bolometric luminosity of the central engine. This allows us to
constrain the mass accretion rate $\dot{M}$ of the black hole based
on the different accretion models which have been
proposed. 

\section{Observations and Results}

The 1.1~$\micron$ to 2.2~$\micron$ data were obtained using the Near Infrared Camera and
Multi-Object Spectrometer (NICMOS) Camera 2 on the Hubble Space Telescope
(Thompson et al. 1998). The camera has a 256$\times$256 HgCdTe array with
pixel scales of 0.0762$\arcsec$ and 0.0755$\arcsec$ per pixel in $x$ and
$y$, respectively, providing a $\sim19.5\arcsec\times19.3\arcsec$ field of
view.  Frames were taken in three filters: F110W, F160W and
F222M, which correspond to 1.1~$\mu$m, 1.6~$\mu$m, and 2.2~$\mu$m 
respectively.  Observations of the galaxy were made using a four-point
spiral dither in each filter setting; the offset per dither position was
25.5 pixels (1.9$\arcsec$). Non-destructive reads (MULTIACCUM) were
obtained at each dither position, with integration times of 40 (F110W,
F160W) and 56 (F222M) seconds per position. The total
integration times were thus 160 (F110W, F160W) and
224 (F222M) seconds.  In addition to the galaxy observations, blank-sky
observations were done with the F222M filter to determine the background
level at 2.2~$\mu$m, and dark frames were taken using the same MULTIACCUM 
sequences as the galaxy observations.

The data were reduced with IRAF. The dark was first created, then
the NICMOS data were dark subtracted, flatfielded and corrected for cosmic
rays using the IRAF pipeline reduction routine CALNICA (Bushouse 1997; See
Scoville et al. 1999 for details about the reduction procedure).
The calibrated images contained pixels with reduced quantum efficiency due
to contaminants on the array surface, thus a mask was created to minimized
their effect.  The coronographic hole and column 128 were also masked on
all images, the latter feature was masked due to its sensitivity to minute
discrepancies in dark subtraction (i.e., the dark current rises sharply
toward the center column of the array).  The dithered images were then
shifted and averaged using the DRIZZLE routine in IRAF (Fruchter \& Hook 1997). 
Finally, a background level of 0.45 ADU/s was measured
from the blank-sky data, and this amount was subtracted from the 2.2~$\mu$m
image of NGC 4258. The background level at 1.1~$\mu$m and 1.6~$\mu$m was 
ignored since it is typically less than 0.01 ADU/s.
The plate scales of the final `drizzled' images, shown in Figure 1, are
0.0381$\arcsec$ and 0.0378$\arcsec$ per pixel in $x$ and $y$.

Flux calibration of the images were done using scaling factors of
2.195$\times10^{-6}$, 2.207$\times10^{-6}$,
and 5.583$\times10^{-6}$ Jy~(ADU/sec)$^{-1}$ at
1.1, 1.6, and 2.2 $\mu$m (Rieke et al.~1999). The
corresponding magnitudes were calculated using the zero-points
1909, 1087 and 665 Jy at 0 mag for 1.1, 1.6 and
2.2 $\mu$m respectively.
The 1.1~$\micron$, 1.6~$\micron$ and 2.2~$\micron$ images were aligned to within
0.2 pixel using centroids and elliptical isophotes generated. 
Synthetic NICMOS point spread functions (PSFs) were generated
using the TINYTIM routine (Krist \& Hook 1997) since no
observed PSFs were available. 

As shown by CB97, the nucleus of the galaxy is clearly redder than the
surrounding starlight. The colors of the
nuclear regions (\ab0.12$\arcsec$ radius from the center) 
are [1.6]$-$[2.2]=0.9$\pm$0.1~mag and [1.1]$-$[1.6]=1.1$\pm$0.1~mag.
In comparison, the colors
of the galaxy in an annulus of radius 0.5$\arcsec$ to 1.0$\arcsec$ are
[1.6]$-$[2.2]=0.52$\pm$0.05~mag and [1.1]$-$[1.6]=0.91$\pm$0.05~mag.
The reddening can be attributed to either 
a red central source or dust extinction of starlight. CB97 argued
for the former and we provide additional evidence in this paper that this
is true. 
To estimate the contribution of the putative source seen by CB97
at the nucleus of the
galaxy, it was necessary to subtract a stellar background from the galaxy
frames. The 1.1~$\micron$ surface brightness profile 
was fitted with an exponential and subtracted 
off such that
the residual consisted of a central core with a FWHM corresponding to
the FWHM of the 1.1~$\micron$ PSF (\ab0.1$\arcsec$). There is no evidence of an Airy
ring in the residual and hence we conclude that we do not detect a
compact component of emission at 1.1~$\micron$. The residual 
has a flux density of 0.5 mJy and we consider it as an upper limit
to a point source. 

Since the 1.1~$\micron$ data do not show a compact source, it is
reasonable to assume 
that it traces the distribution of starlight in the central regions 
of the galaxy. Furthermore, the contribution from the polarized
component observed by W95, based on it's
flux density and spectrum, is found to be less than 0.1 mJy
at 1.1~$\micron$.
A scaled 
1.1~$\micron$ surface brightness image was subtracted
from the 1.6~$\micron$ and
2.2~$\micron$ surface brightness images. The scale factor
was derived by performing an intensity-weighted
least squares fit within the central
arcsecond of the form:
$\Sigma_{[2.2]}$=a$_{1}$~$\Sigma_{[1.1]}$ + b$_{1}$~P$_{[2.2]}$ where
$\Sigma$ is the surface brightness, a$_{1}$ and b$_{1}$ are constants 
and P is the TINYTIM generated PSF.
For the 2.2~$\micron$ image, we obtain a$_{1}$=1.25$\pm$0.05 and for the
1.6~$\micron$ image, we get a$_{1}$=1.29$\pm$0.05. 
This  corresponds to [1.1]$-$[2.2] and
[1.1]$-$[1.6] colors of 1.39$\pm$0.05 mag and 0.89$\pm$0.05 mag 
respectively. 

The Airy ring is clearly visible in the starlight subtracted 1.6~$\micron$ and
2.2~$\micron$ images which were obtained as described above. 
The FWHM of the residual is also very similar to the FWHM of the
synthetic PSFs at the respective wavelengths.
We thereby conclude that the compact component of emission 
at 1.6~$\micron$ and 2.2~$\micron$ is unresolved at our resolution of \ab0.2$\arcsec$.
This implies that the size of the emitting region has a radius $\lesssim$3.5 pc 
at the distance of the galaxy and is consistent with the
emission originating from the vicinity of the black hole at the nucleus.
The profile of the central source is shown in Figure 2.
The brightness of the central source was measured by standard
circular aperture photometry. 
The observed flux density of the compact source is 4.0$\pm$0.7
mJy at 2.2~$\micron$ and 0.9$\pm$0.3 mJy at 1.6~$\micron$.
This agrees well with the values
of 4.5~mJy and 1.1~mJy obtained by CB97 at \K~and \H~respectively.

Rieke (1999) has suggested that using the standard NICMOS calibration numbers 
for red objects
as above, could result in an error of as much as
0.7 mags in the intrinsic NICMOS magnitude especially at [1.1]. The effect
is smaller at [1.6] and negligible at [2.2]. This is because red objects 
would shift the effective wavelength and
the zero point of the F110W and F160W filters, both of which have passbands
much wider than the ground-based $J$ and $H$ filters.
We estimate the amount of this error for NGC4258
by applying a conversion from
instrumental NICMOS magnitudes to ground-based magnitudes (Appendix A).
This also serves a secondary purpose which is
to compare the colors of the starlight in NGC4258, derived
from the NICMOS data, with ground-based colors.

We have converted the NICMOS [1.1] and [1.6] images to ground-based
$J$ and $H$ magnitudes and rederived the compact source flux density
using the procedure described earlier. We find that the starlight
in the galaxy has a \J$-$\K~color of 1.0 mag and a \J$-$\H~color
of 0.6 mag. This is in excellent agreement with the 
ground-based colors of
\J$-$\K$=$0.9 mag and \J$-$\H$=$0.6 mag obtained by CB97
at a distance of 6$\arcsec$ from the center.
These colors are consistent with the colors
of late
type stars expected to dominate the surface brightness in the central regions
of the galaxy.
The resultant flux density of the central source is 1.1 mJy at 1.6~$\micron$
and 3.6 mJy at 2.2~$\micron$. This is within the photometric errors assigned to the
compact component above and hence we conclude that the wide bandpass
of the filters
doesn't significantly affect our determination of the compact source flux density.

Near infrared data 
were also obtained in the \Ls~($\lambda_{0}=3.45~\micron$, $\Delta\lambda=0.57~\micron$) 
and \Ks~($\lambda_{0}=2.15~\micron$, $\Delta\lambda=0.33~\micron$)
bands
using the Cassegrain IR Camera on the 5.1 m Hale 
telescope. A \Ks~PSF source was scaled to the 
flux density of the compact 2.2~$\micron$ source in NGC4258 and subtracted
from the \Ks~image of the galaxy after aligning.
The residual image corresponds to the surface brightness
profile of the starlight in the galaxy. 
This was then normalized to the \Ls~surface brightness 
at a radius of 2.25$\arcsec$ and subtracted. The normalization
factor of $\Sigma_{Ks}/\Sigma_{Ls}$=1.98 corresponds to 
a \Ks$-$\Ls~color of 0.16 mag, which is reasonable
considering the \J$-$\H~and \H$-$\K~colors of the starlight. 
Photometry was performed on the
\Ls~residual image which is dominated by the compact component.
We measure a flux density of 20$\pm$3 mJy for this source.

The mid infrared data 
were obtained using the MIRLIN camera on the 10 meter Keck II 
telescope in March 1998 (Ressler et al. 1994). 
The array is a 128$\times$128 array with a pixel scale of
\ab0.14$\arcsec$/pixel at the f/40 bent Cassegrain focus of Keck II. 
The instrument has a square wave
chopper which chops at 5 Hz. The chopper throw was set at 6$\arcsec$
which allowed the source to be on the array at both chop
positions. Observations were made with a filter which is centered 
at 12.5~$\micron$ with a passband of
1.2~$\micron$ and a filter which is centered at 17.9~$\micron$ with a
passband of 2~$\micron$. We chopped 300 times per position and
nodded by 6$\arcsec$ after a minute. 
The total integration time was
300 s at 12.5~$\micron$ and 450 s at 17.9~$\micron$. The resultant
1$\sigma$ sensitivity at 12.5~$\micron$ and 17.9~$\micron$ was 20 mJy 
and 50 mJy in a 2.8$\arcsec$ diameter beam respectively. 

The mid infrared data were reduced in a standard manner (Bock et al.
1998).
Airmass corrections were performed which were less than 5$\%$ since the
standards and source were at similar elevation angles. The final image
at these wavelengths is dominated by a single unresolved 
source (FWHM$\lesssim$0.5$\arcsec$) with no evidence of
extended emission on scales of diameter up to 6$\arcsec$. 
The observed flux density of the unresolved
central source is 165$\pm$20 mJy at
12.5~$\micron$ and 300$\pm$30 mJy at 17.9~$\micron$.

\section{Emission from a Compact Extincted Continuum Source}

CB97 showed that the red near-infrared colors of the nucleus of the
galaxy could not be explained by uniform foreground extinction of starlight
or by stars mixed with dust. In addition, starlight cannot explain the
observed spectrum of the enhanced nuclear emission 
at $\lambda > 3~\micron$ (Table 1).

If the enhanced infrared emission in the nucleus arises in
a compact nuclear continuum source associated with the 
3.5$\times$10$^{7}$~M$_{\sun}$ black hole as argued by CB97, 
it implies that the extinction 
observed in the X-ray data occurs in
a foreground screen such as a dust lane, attenuating both the near
and mid infrared intensities\footnote{For the rest
of this paper, we assume A$_{[1.1]}$=0.282~A$_{V}$,
A$_{[1.6]}$=0.624~A$_{[1.1]}$,
A$_{[2.2]}$=0.382~A$_{[1.1]}$, A$_{[3.5]}$=0.182~A$_{[1.1]}$,
A$_{[10]}$=0.192~A$_{[1.1]}$,
A$_{[12.5]}$=0.098~A$_{[1.1]}$,
A$_{[17.9]}$=0.083~A$_{[1.1]}$ (Mathis 1990).}.
The visible/UV polarimetry data on the central source
shows a F$_{\nu}\propto\nu^{-1.1\pm0.2}$ spectrum in scattered light 
(W95) similar to our observed mid-infrared spectrum.
Therefore, it is reasonable to assume that the intrinsic near-infrared 
spectrum of the compact source follows
a similar trend and that the observed [1.6]$-$[2.2] color 
of the compact source is due to extinction. 
We calculate that the amount of extinction required to fit the 
near and mid infrared data by a single power law 
corresponds to A$_{V}=$18 mag. This is smaller than the extinction
derived from the X-ray column density. 
However, as pointed out by Gammie, Narayan, \& Blandford (1999),
the column density N$_{H}$ derived by Makishima et al. (1994) is 
unreliable since it is sensitive to the relative contributions of the
components used to fit the X-ray data. In addition,
the conversions from the X-ray opacity, $\tau_{X}$ to N$_{H}$ and from N$_{H}$
to A$_{V}$ is not well established.
The resultant extinction corrected infrared spectrum 
follows a $\nu^{-1.4\pm0.1}$ power law, suggesting a non-thermal origin. 
The extinction corrected flux densities are shown in Figure 3 and Table 1.

The extrapolated flux density of the central source at 1.1~$\micron$
based on our fitted spectrum
is calculated to be 10 mJy. If we extinct this 1.1~$\micron$ source 
by A$_{V}$=18~mag, we obtain a flux density
of 0.1 mJy, consistent with our derived upper
limit. The 0.6$-$20~$\micron$ luminosity of 
the central source derived from our fitted F$_{\nu}\propto\nu^{-1.4}$ spectrum 
is 8.5$\times$10$^{41}$ ergs s$^{-1}$. This is a factor of 20 greater 
than the X-ray luminosity. 

RL78 measured
a flux density of 100$\pm$12 mJy in a 5.7$\arcsec$ diameter beam with a 8$-$13~$\micron$ filter.
Since our mid-infrared images do not show any evidence of extended emission
from the nucleus of the galaxy at scales of 6$\arcsec$, 
this measurement must be dominated by the central source.
If our extinction corrected measurements are interpolated to 10~$\micron$ assuming a
$\nu^{-1.4}$ spectrum, we obtain an unextincted flux density 
in the 8$-$13~$\micron$ filter of 190 mJy. 
After putting in the extinction in the 8$-$13~$\micron$ window
including the 9.7~$\micron$ silicate feature, corresponding to A$_{V}$=18
mag, we derive an extincted 10~$\micron$ flux density of 98~mJy for the central
source, in excellent agreement with the large beam measurements. 

The VLA observations of Turner \& Ho (1994) had detected the
existence of a non-thermal source in the nucleus of the galaxy at 2 cm
and 6 cm. Milli-arcsecond resolution VLBI continuum observations indicate
that most of the non-thermal radio emission arises in a sub-parsec-scale jet which is
aligned with the rotation axis of the masing disk (Herrnstein et al.
1997). The black hole region itself is undetected and Herrnstein et
al. (1998) obtain a 3$\sigma$ upper limit of 220~$\mu$Jy at 22 GHz.
When compared to the extinction corrected infrared spectrum, this seems 
to suggest that the spectrum has a turnover probably due to
synchrotron self-absorption between
20~$\micron$ and 0.2 mm. The total luminosity of the central source longward of
1~$\micron$ is relatively insensitive to the turnover 
frequency and is constrained to lie between
8$\times$10$^{41}$ ergs s$^{-1}$ and 3$\times$10$^{42}$ ergs s$^{-1}$
for turnover wavelengths of 20~$\micron$ and 0.2 mm respectively.
Spinoglio \& Malkan (1989) suggested that the 12~$\micron$ luminosity
is typically about 1/5 of the bolometric luminosity of active galaxies,
independent of whether the emission is thermal or non-thermal in origin. 
We find by this argument, the
bolometric luminosity of NGC4258 to be \ab10$^{42}$ ergs s$^{-1}$,
which is in between our independently derived luminosity range.

As indicated by the extinction correction to our measurements, the
nucleus of NGC4258 clearly has dust associated with it.
The extinction corrected infrared emission 
from 1~$\micron$ to 20~$\micron$ 
has a $\nu^{-1.4}$ power law spectrum which
appears to extend up to visible wavelengths as seen in
the polarimetric observations of W95. This makes thermal dust emission
at these wavelengths unlikely and we infer the infrared emission is 
non-thermal in origin. 
Many previous authors have concluded that the source of both the near and 
mid infrared
emission in the nucleus of Seyferts is thermal emission from dust 
reprocessing of a blue, nuclear continuum (e.g. Giuricin, Mardirossian, \& Mezzetti 1995). 
This was  
based on a weak correlation between the X-ray and
mid-infrared emission in a large sample of Seyferts. 
Our derived spectrum suggests that
it is more likely that the dust in the nucleus of NGC4258 is cool and 
contributes substantially only at
far-infrared/submillimeter wavelengths.

\section{The Accretion Flow in NGC4258}

\subsection{Implications for an Advection Dominated Flow}

M97 has derived the 10$^{8}$ to 10$^{20}$ Hz spectrum 
of systems dominated by an ADAF (Figure 4).
The spectrum consists of three components; a synchrotron component
at $\nu<10^{12}$ Hz, a power law component for 10$^{12} < \nu < 10^{17}$ Hz
from the inverse Compton scattering of the cyclosynchrotron photons
and a bremsstrahlung component at higher frequencies.

From M97,
the spectral index of the power law (L$_{\nu}\propto \nu^{-\gamma}$) is given by
\begin{equation}
\gamma\equiv\frac{-\ln \tau_{es}}{\ln A}
\end{equation}
\begin{equation}
\tau_{es}=(23.87~\dot{m})\left(\frac{0.3}{\alpha}\right) \left(\frac{0.5}{c_{1}}\right) \left(\frac{3}{
r_{min}}\right)^{1/2}
\end{equation}
\begin{equation}
A = 1 + 4\theta_{e} + 16\theta_{e}^{2}
\end{equation}
\noindent where $\tau_{es}$ is the optical depth to electron scattering,
A is the mean photon energy amplification factor in a single scattering,
$\theta_{e}=(k T_{e})/(m_{e}c^{2})$, $c_{1}=0.5$ is a term
relating the radial velocity of the accretion flow to the Keplerian
velocity
and $r_{min}$ is the inner radius of the ADAF which we assume to be 3~R$_{S}$.
$T_{e}$ is the electron temperature which is constant over the
ADAF region.
Based on the figures from M97, we adopt a $T_{e}$=4.4$\times$10$^{9}$~K,
$\alpha$=0.3, and derive from the above equations,
an accretion rate of $\dot{m}\sim1.2\times10^{-3}$ in Eddington units,
which corresponds to $\dot{M}\sim9\times10^{-4}$ M$_{\sun}$/yr.

It is also possible to estimate $\dot{m}$ from the
bolometric luminosity as shown by M97. For $\alpha$=0.3,
$\beta$=0.5, we find that our L$_{Bol}\sim10^{42}$ ergs~s$^{-1}$
corresponds to $\dot{m}\sim6\times10^{-3}$ in Eddington units which
is again reasonable considering the factor of 3 uncertainty in
the bolometric luminosity. $\beta$ is the ratio of gas pressure to 
total pressure.

In comparison, Lasota et al. (1996) derived a rate of $\dot{m}\sim0.16\alpha$
through an ADAF.
However, at that time the infrared data
which better constrain
the bolometric luminosity of the central engine was not available and the ADAF
models were fit to the slope of the X-ray data. More recently, Gammie et al.
(1999) derive an accretion rate of 0.01 M$_{\sun}$/yr through an ADAF 
using the infrared data of CB97 as a constraint. 
Again, this analysis assumed that most
of the visible/IR emission has a $\nu^{1/3}$ Comptonized blackbody spectrum
and
arises in the cool, thin disk rather than in the ADAF
itself. Our recent data
suggests that the visible/IR emission has a common origin
and has an intrinsic spectrum that can be well represented by a
\ab$\nu^{-1}$ power law. 

Herrnstein et al. (1998) have shown based on their 22 GHz
upper limit, that the ADAF cannot extend significantly beyond
\ab10$^{2}$ R$_{S}$. We find that the 22 GHz limit 
in conjunction with the spectrum of M97 in the
self-absorbed radio regime is inconsistent 
with our measurements (Figure 4). We predict that the radio synchrotron spectrum must
have a spectral index of 2.5$\geq \acute{\gamma} \geq 1.2$ 
where F$_{\nu}\propto\nu^{\acute{\gamma}}$ rather than 
the $\acute{\gamma}=\frac{2}{5}$
derived for a thermal distribution of electrons by M97. 
This could arise if the distribution of electrons has a power law
tail at higher energies. 
Maintaining a power law distribution of relativistic particles
without resorting to an independent injection mechanism such as
magnetic reconnection has been an unsolved problem for
some time now (McCray 1969).
All previous work has assumed that the particle distribution in an ADAF
is thermal. However, 
Mahadevan \& Quataert (1997) investigated the emission spectrum for
different accretion rates through an ADAF. They found that for low accretion
rates, the thermalization of the electron distribution through
Coulomb collisions with the protons and through self absorption of synchrotron
photons, is inefficient. Adiabatic compression of the electron gas acts
as the primary heating mechanism which results in 
the high energy tail of the electron distribution function 
having a non-thermal Gaussian profile rather than a Maxwellian profile.
Alternatively, Mahadevan et al. (1997) has suggested that if the
weak Galactic center source Sgr A$^{*}$ is
associated with the EGRET source 2EG J1746-2852, the gamma ray
spectrum could be explained by an E$^{-2.75}$ distribution
of protons in an ADAF.
A similar power law distribution of electrons would indicate that
the electron distribution function is generated by the
same viscous heating mechanism responsible for generating the
proton distribution function.

\subsection{Implications for a Standard Thin Disk}

We find that the excellent agreement 
between the fit F$_{\nu}\propto\nu^{-1.4}$ spectrum and data obviates the need
for a significant $\nu^{\frac{1}{3}}$ Comptonized blackbody
continuum from a standard optically thick, geometrically thin disk.
However, 22 GHz observations
show the existence of a thin disk with inner radius 0.14 pc which sustains the
water masers (Miyoshi et al. 1995). From standard disk theory (Frank, King \& Raine 1992),
\begin{equation}
T_{eff}\sim3.5\times10^{7}~m^{-1/4}~\dot{m}^{1/4}~r^{-3/4}~{\rm K}
\end{equation}
\noindent where $m$ is the central mass in units of M$_{\sun}$ 
and $r$ is the radius of the disk in units of R$_{S}$.
For a geometrically thin,
optically thick disk to be capable of emitting blackbody near-infrared photons, it
should be at least 1500~K. Based on the derived accretion rate, this
implies a conservative lower limit of 200~R$_{S}$ for the
inner radius of the thin disk.

\section{Conclusions}

NGC4258 has associated with it a compact nuclear source at
near and mid infrared wavelengths which is unresolved even 
at a near-infrared resolution of \ab0.2$\arcsec$ FWHM. This implies the emitting
region is $\lesssim$3.5~pc in radius at the distance of the galaxy.
The observed infrared flux densities can be fit
by a single power law of the 
form F$_{\nu}\propto\nu^{-1.4\pm0.1}$ from 1~$\micron$ to 18~$\micron$
after correcting for
about 18 mags of visual extinction. 
The total extinction corrected
infrared luminosity (1$-$20~$\micron$) of the central source is
8$\times10^{41}$ ergs s$^{-1}$.
At shorter wavelengths,
the power law spectrum seems to extend upto visible wavelengths as 
shown by the
polarimetric observations of W95. 
At longer wavelengths, the infrared spectrum along with the 22 GHz upper limit
points towards a
synchrotron self-absorption turnover wavelength between 20~$\micron$ and 0.2 mm.
From this spectrum,
we derive the bolometric luminosity of the nuclear source to be
\ab2$\times$10$^{42}$~ergs~s$^{-1}$.
The low bolometric luminosity and power law spectrum extending from visible
to mid-infrared wavelengths indicates that
self-Comptonized synchrotron emission arising in an ADAF is
responsible for the enhanced infrared emission.
The spectrum obviates the need for a ``big blue bump" at
these wavelengths from a cool, optically thick,
geometrically thin disk extending beyond the advection dominated flow. 
Such a disk is predicted to have a F$_{\nu}\propto\nu^{1/3}$
spectrum which appears unlikely from our data.
Based on the spectrum of the source and the derived bolometric luminosity,
we infer the accretion rate through an ADAF in NGC4258 to 
be \ab10$^{-3}$~M$_{\sun}$/yr.

\acknowledgements

This work is based on observations made at the W. M. Keck 
Observatory, Palomar Observatory and with the 
NASA/ESA {\it Hubble Space Telescope}, which is
operated by the Association of Universities for Research in 
Astronomy, Inc. under NASA contract No. NAS5-26555. 
We wish to acknowledge the contributions made by 
members of the NICMOS Instrument Definition Team to this project. 
We are grateful to Steve Willner for his help with clarifying 
the discussion in the Appendix.
We would also like to thank the staffs of 
the respective observatories for their assistance.
The W. M. Keck Observatory is a scientific partnership between the
University of California and the California Institute of Technology. 

\appendix

\section{Conversion of NICMOS magnitudes to Ground-based magnitudes}

In this section, we describe the color correction applied to convert
the NICMOS [1.1], [1.6] and [2.2] magnitudes to ground-based $J$, $H$, $K$ 
magnitudes. The color correction
is applied for two reasons:\\
1. To compare the results from this NICMOS data set with previous
ground-based observations,\\
2. To correct for calibration errors which arise due to the wider
bandpass of the NICMOS F110W and F160W filters. 
It has been suggested that since the NICMOS filters were 
calibrated off solar analog stars,
using the standard NICMOS calibration numbers could result in an
error of as much as 0.7 mags in the intrinsic NICMOS
magnitude for red ($J-H>1$~mag) objects (Rieke 1999). This error is not significant in the
ground-based $J$, $H$ filters because of their much narrower bandpass.

NICMOS has observed a sample of standards of different colors for
which the ground-based magnitudes are known (Table 2). Thus, by converting
instrumental magnitudes
to ground-based magnitudes on the basis of these observations, we can eliminate
calibration errors.
Figure 5 shows the color correction with the best corresponding linear fit 
listed below. The color correction at [2.2] is small enough that we take it to be 0.
\begin{eqnarray}
J-[1.1] &=& - 0.58\times([1.1]-[1.6]) + 0.11\\
H-[1.6] &=& -0.24\times([1.6]-[2.2]) + 0.01
\end{eqnarray}

\clearpage

\begin{figure}
\plotone{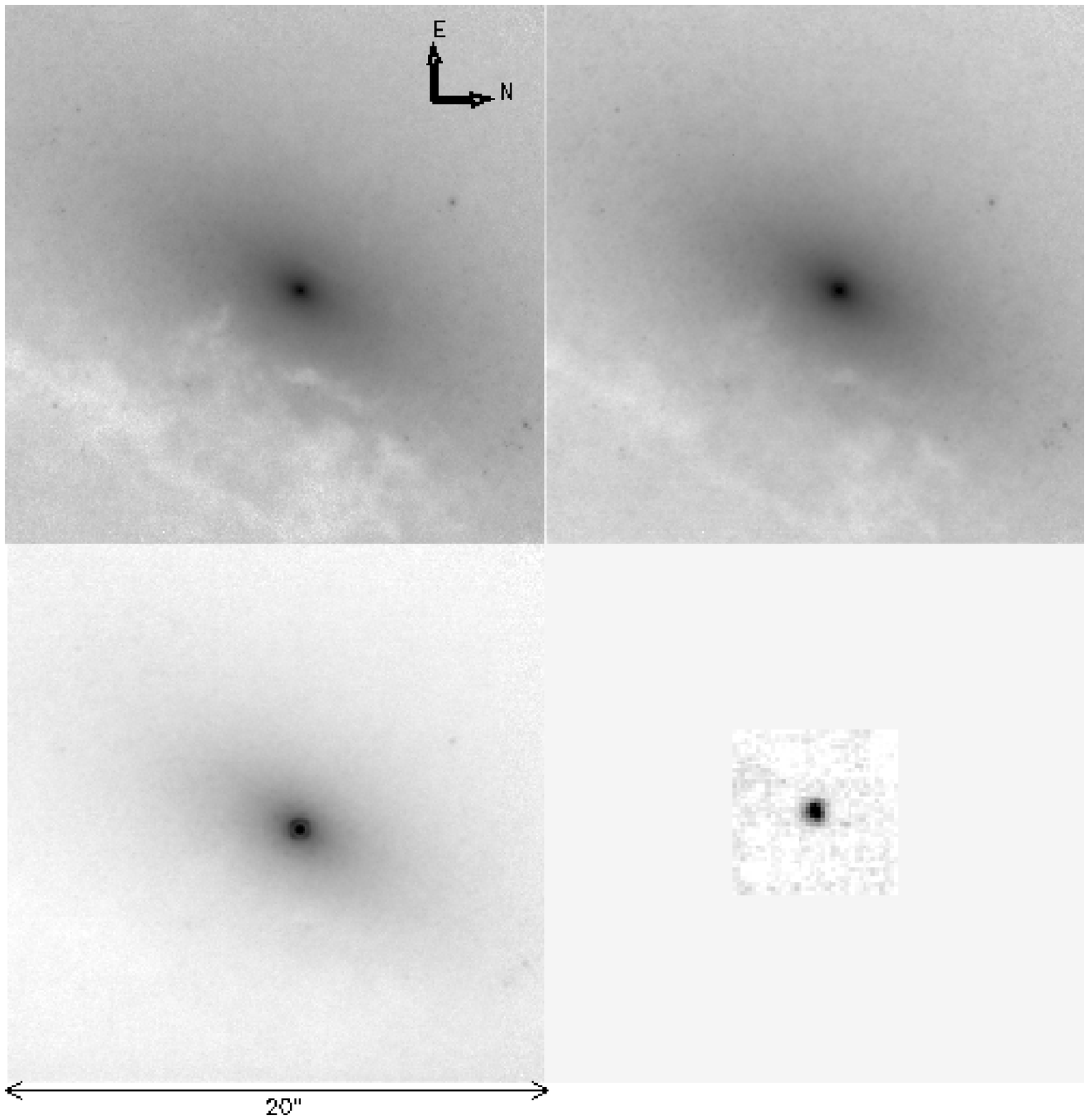}
\caption{Clockwise from top left: Morphology of the nuclear regions in
NGC4258 at 1.1~$\micron$, 1.6~$\micron$, 
12.5~$\micron$ and 2.2~$\micron$. The near-infrared images are 20$\arcsec$ on a side. 
The 1.1~$\micron$ and 1.6~$\micron$ images clearly show
the dust lane in the nuclear regions of the galaxy. 
The 12.5~$\micron$ image
which is 6$\arcsec$ on a side does not show any extended emission.
$\alpha=12h~18m~57.5s, \delta=47\arcdeg~18'~14''$ (J2000).}
\end{figure}

\begin{figure}
\plotone{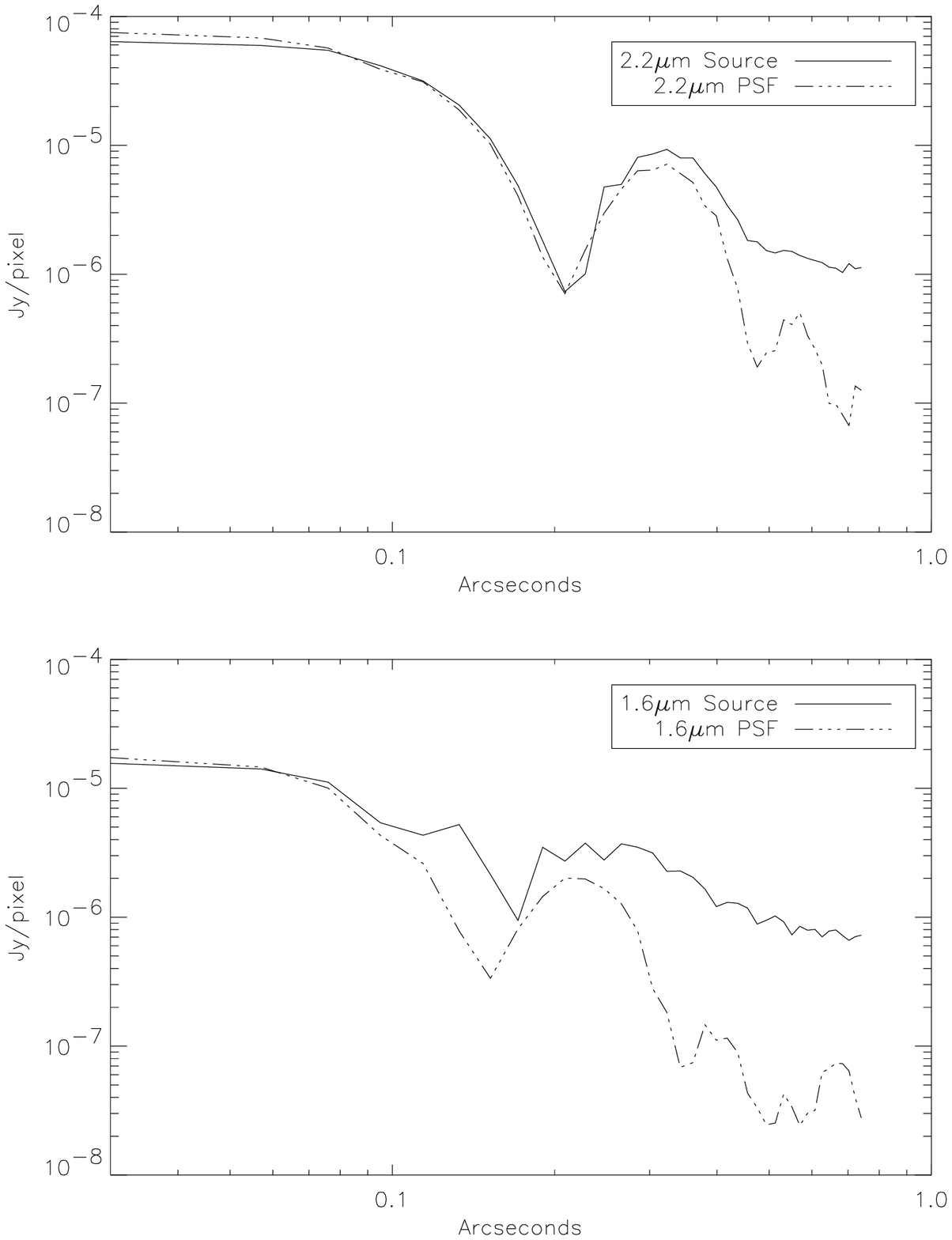}
\caption{The profile of the central source at 2.2~$\micron$ and
1.6~$\micron$ compared to a TINYTIM generated point source.}
\end{figure}

\begin{figure}
\plotone{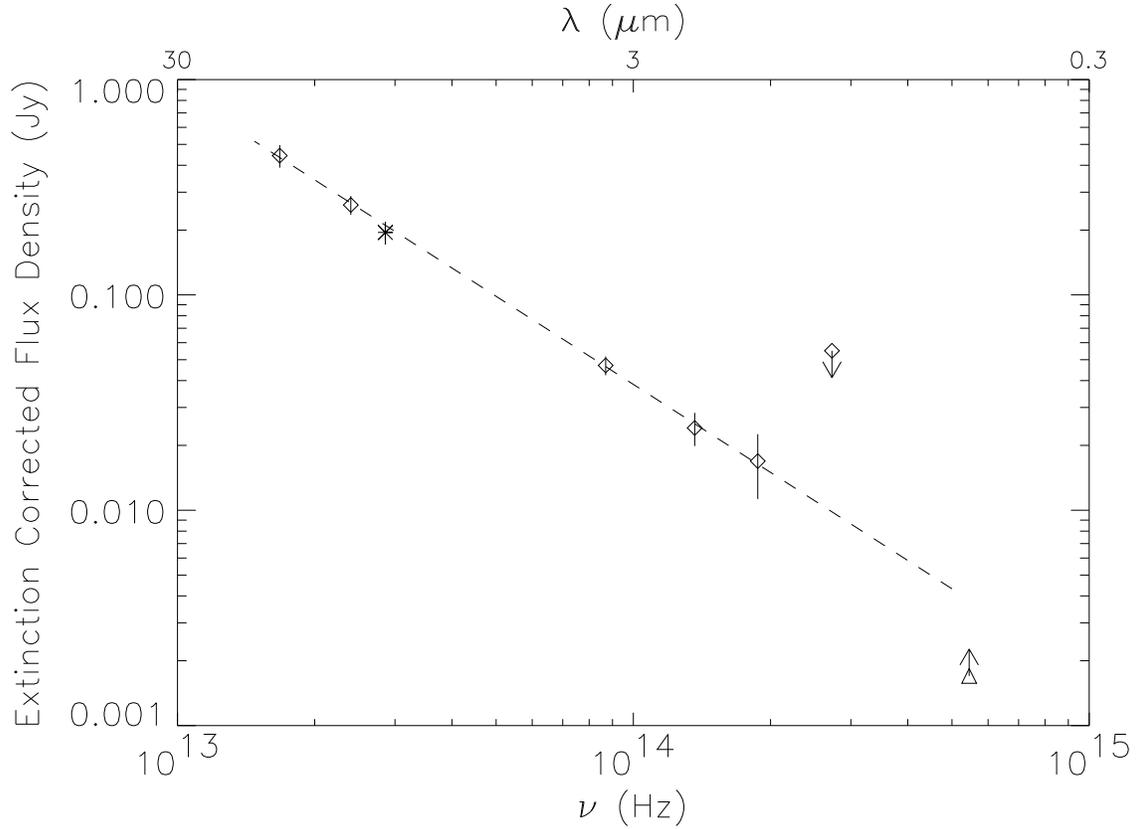}
\caption{The plot shows the fitted spectrum (F$_{\nu}\propto\nu^{-1.4}$)
of the central source along with the near and mid infrared data. 
These assume foreground extinction
by a screen so both the near and mid infrared points have been
extinction corrected. The
visible light polarimetry measurement (triangle) of W95 
is shown as a 
lower limit because of the large uncertainty in the value. The
asterisk represents the extinction corrected RL78 measurement.}
\end{figure}

\begin{figure}
\plotone{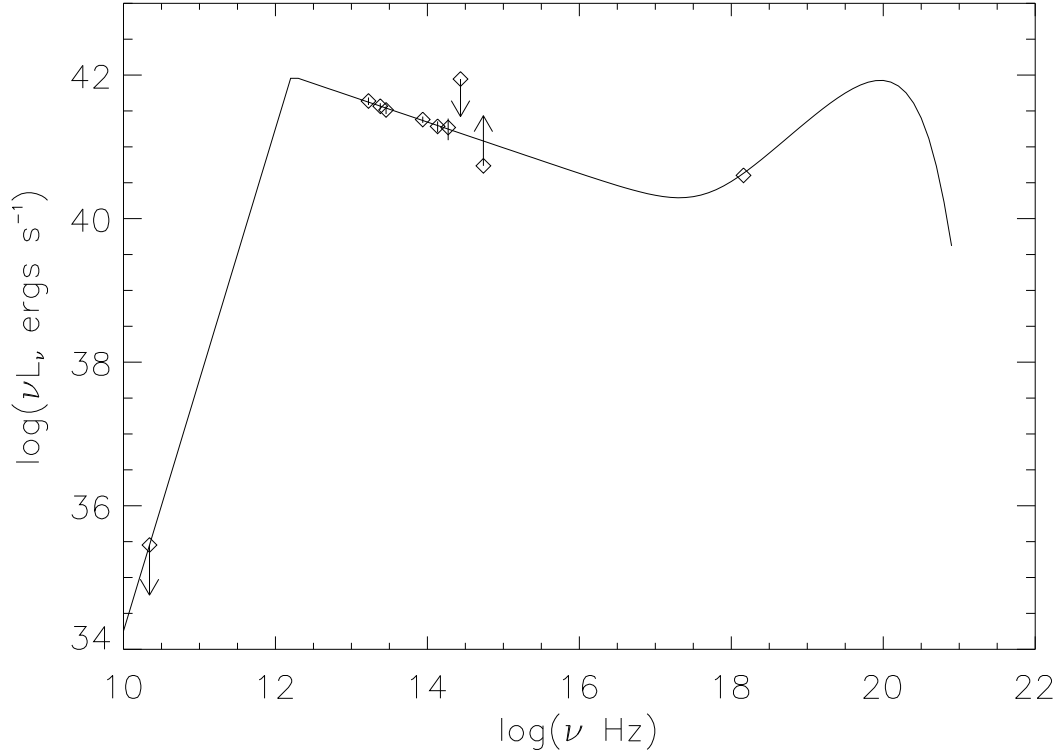}
\caption{Spectrum of an ADAF fit to the data on NGC4258.
There are three components of emission, a L$_{\nu}\propto\nu^{5/2}$ synchrotron
component, a L$_{\nu}\propto\nu^{-1.4}$ power law component from Compton upscatter of
the synchrotron photons and a bremsstrahlung component with an exponential cutoff
at frequencies h$\nu\geq k T_{e}$. The components from $\nu > 10^{12}$~Hz agree with
M97. Below that we have adopted a $\nu^{5/2}$ synchrotron component instead of the
$\nu^{2/5}$ predicted in M97. The W95
visible light polarimetry is shown as a lower limit because of the 
large uncertainty in the measurement. The plot also shows the 1.1~$\micron$
upper limit derived from our observations, the 22 GHz upper limit
of Herrnstein et al. (1998) and the Makishima et al. (1994) X-ray measurement.}
\end{figure}

\begin{figure}
\plotone{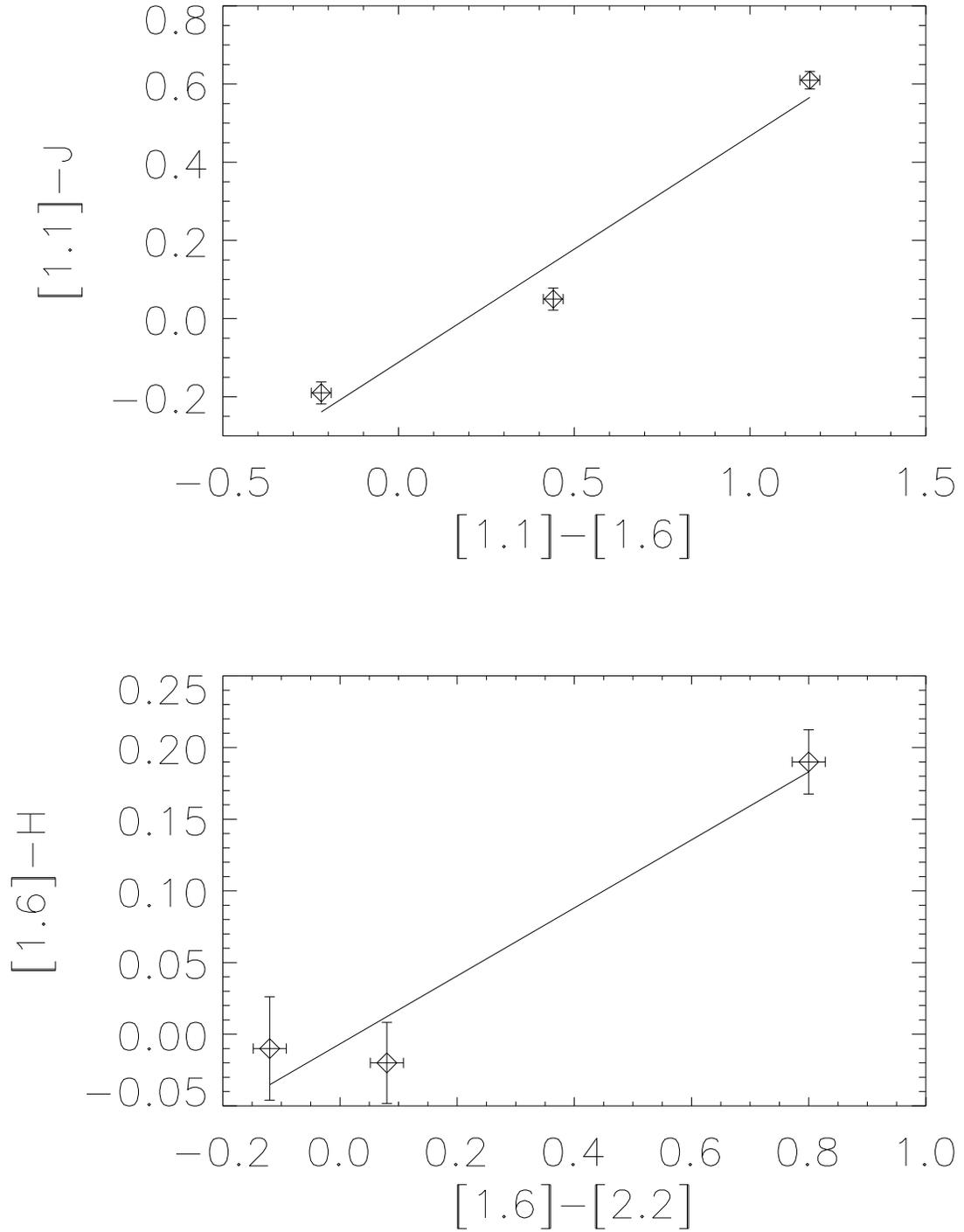}
\caption{Color correction between NICMOS magnitudes and ground-based magnitudes based on
observations of standards of different colors. The lines are the best fit first-order polynomials
described in the text. The color correction at [2.2] is small enough that we take it to be 0.}
\end{figure}

\begin{deluxetable}{ccccc}
\singlespace
\tablecaption{Energy Distribution for the Compact Source}
\tablehead{
\colhead{$\lambda$} &
\colhead{$\nu$} &
\colhead{Observed F$_{\nu}$} &
\colhead{Extinction Corrected F$_{\nu}$} &
\colhead{Reference}\\
\colhead{} &
\colhead{Hz} &
\colhead{mJy} &
\colhead{mJy} &
\colhead{}
}
\tablewidth{0pt}
\tablecolumns{5}
\startdata
2$-$10~keV&1.45$\times10^{18}$&...&4.7$\times10^{-4}$& Makishima et al. (1994)\\
0.55$\micron$&5.45$\times10^{14}$&...&1.7$-$170&W95\\
1.1$\micron$&2.73$\times10^{14}$&$<$0.5&10\\
1.6$\micron$&1.88$\times10^{14}$&0.9$\pm$0.3&16\\
2.2$\micron$&1.36$\times10^{14}$&4.0$\pm$0.7&25\\
3.45$\micron$&8.70$\times10^{13}$&20$\pm$3&46\\
10.5$\micron$\tablenotemark{a}&2.86$\times10^{13}$&100$\pm$12&195&RL78\\
12.5$\micron$&2.40$\times10^{13}$&165$\pm$20&270\\
17.9$\micron$&1.68$\times10^{13}$&300$\pm$30&435\\
1.4 cm&22$\times10^{9}$&...&$<$0.22& Herrnstein et al. (1998)\\ 
\enddata
\tablenotetext{a}{8$-$13~$\micron$ filter.}
\end{deluxetable}

\begin{deluxetable}{ccccccc}
\singlespace
\tablecaption{Photometry of Unextincted NICMOS Stars\tablenotemark{b}}
\tablehead{
\colhead{Object} &
\colhead{[1.1]} &
\colhead{$J$} &
\colhead{[1.6]} &
\colhead{$H$} &
\colhead{[2.2]} &
\colhead{$K$} \\
}
\tablewidth{0pt}
\tablecolumns{7}
\startdata
G191-B2B (WD) & 12.49$\pm$0.02 & 12.68$\pm$0.02 & 12.71$\pm$0.02 & 12.72$\pm$0.03 & 12.83$\pm$0.02 & ... \\
P330E (G star) & 12.01$\pm$0.02 & 11.96$\pm$0.02 & 11.57$\pm$0.02 & 11.59$\pm$0.02 & 11.49$\pm$0.02 & 11.51$\pm$0.02 \\
BRI0021 (red) & 12.45$\pm$0.02 & 11.84$\pm$0.01 & 11.28$\pm$0.02 & 11.09$\pm$0.01 & 10.48$\pm$0.02 & 10.55$\pm$0.01 \\
\enddata
\tablenotetext{b}{Obtained from the STScI wesite with
the errors not representing calibration errors.}
\end{deluxetable}

\end{document}